\documentclass[oneside,english]{aa}
\usepackage[T1]{fontenc}
\usepackage[latin1]{inputenc}
\usepackage{babel}
\makeatletter

\usepackage{txfonts}

\usepackage{natbib}
\bibpunct{(}{)}{\ and}{a}{}{,}

\usepackage{graphicx}
\usepackage{amstext}
\usepackage{amscd}
\usepackage{bbm}

\newcommand{\Ableitung}[2]{\frac{\mathrm{d}#1}{\mathrm{d}#2}}

\renewcommand{\text}[1]{\mathrm{#1}}
\newcommand{\R}{\mathbbm{R}}
\renewcommand{\sol}{\odot}

\makeatother
\begin{document}

\title{Numerical stability of mass transfer driven by Roche lobe overflow
in close binaries}

\titlerunning{Numerical stability of mass transfer \ldots}


\author{Andreas B\"uning\and Hans Ritter}

\institute{Max-Planck-Institut f\"ur Astrophysik,
Karl-Schwarzschild-Str. 1, D--85741 Garching, Germany}

\offprints{Hans Ritter \\\email{hsr@mpa-garching.mpg.de}}

\date{Received / Accepted}

\abstract{Numerical computation of the time evolution of the mass
transfer rate in a close binary can be and, in particular, has been a
computational challenge. Using a simple physical model to calculate
the mass transfer rate, we show that for a simple explicit iteration
scheme the mass transfer rate is numerically unstable unless the time
steps are sufficiently small. In general, more sophisticated explicit
algorithms do not provide any significant improvement since this
instability is a direct result of time discretization. For a typical
binary evolution, computation of the mass transfer rate as a smooth
function of time limits the maximum tolerable time step and thereby
sets the minimum total computational effort required for an
evolutionary computation. By methods of ``Controlling Chaos'' it can
be shown that a specific implicit iteration scheme, based on Newton's
method, is the most promising solution for the problem.

\keywords{binaries: close -- stars: evolution -- stars: mass-loss --
methods: numerical -- instabilities}} 

\maketitle


\section{Introduction}
\label{Sec:Introduction}

To compute the long-term evolution of a semi-detached binary, the
numerically determined mass transfer rate, which is a function of 
current stellar and binary parameters, must be a sufficiently smooth
function of time. The simplest approach to obtain the mass $M_2$ of
the mass losing secondary star is an explicit forward integration in
time:
\begin{equation}
\label{Eq:Ansatz:explicit}
M_2(t_{\text {n+1}})=M_2(t_{\text {n}})+\dot{M}_{2}(t_{\text {n}})\, \Delta t.
\end{equation}
Here, $\Delta t = t_{\text{n}+1} - t_{\text{n}}$ denotes the length,
and $t_{\text{n}}$ and $t_{\text{n}+1}$ the start and end points of
the $n$th time step. Every change of $M_2$ within one time step causes
changes of other stellar and binary parameters which affect
$\dot{M}_2$, and an additional change of $M_2$ would be necessary to
obtain a consistent result for the current time step. This feedback
between $\dot{M}_2$ and other stellar and binary parameters,
especially the radius $R_2$ and the critical Roche radius
$R_{\text{R},2}$ of the donor star, can destabilize the numerically
computed mass transfer rate: $\dot{M}_2$ can become non-continuous
between successive time steps and can even show strong fluctuations
around its secular mean value. These numerical effects have been known
for quite a long time (numerical experience by the authors; U. Kolb,
K. Schenker, priv. comm.). Not surprisingly, only very few
evolutionary calculations that show these instabilities of the mass
transfer rate have been published, e.g., \citet[Fig. 3-5]{Ritter90},
\citet[ Fig. 1 \& 9]{Sarna92}, \citet[ Fig. 2]{Antona94},
\citet[ Fig. 3 \& 4]{KR-Cygnus}, and \citet[ Fig. 5]{Schenker02}.

Attempts to use a different explicit integration scheme that takes
into account not only $\dot{M}_2(t_{\text{n}})$ but also
$\dot{M}_2(t_{\text{n-j}})$ for certain values of $j$ (e.g., for a
particular average over the last $m$ time steps) did not show major
improvements. The reason for this behaviour of the mass transfer rate
has remained unknown.

The main purpose of this paper is to show analytically why these
numerical instabilities exist, what they are, which methods are
suitable to suppress them, and which ones are not. Even in the case
of the proposed implicit algorithm, calculation of the mass ransfer
rate still limits the maximum tolerable time step in a numerical
computation and thus sets the minimum total computational effort
required for carrying out such a binary evolution. To illustrate this
point: by imposing a fixed mass loss rate on a single low-mass main
sequence star, up to 10 \% of the total mass of the star can be
removed per time step (eventually after 1 or 2 initial time steps with
a lower rate), and the stellar model still converges. But if the mass 
loss rate is coupled to the binary parameters, even in the case of our
proposed implicit iteration scheme, typically no more than a few
$10^{-3}$ of the stellar mass can be removed per time step without
losing convergence \footnote{%
This upper limit of about $10^{-3}$ varies and depends on the
starting value for the iteration, on the ``smoothness'' of the stellar
input physics, and on the ``smoothness'' of the stellar structure. In
principle, this is not a limitation of the implicit mass transfer
algorithm. 
}. 
When using the explicit algorithm (\ref{Eq:Ansatz:explicit}), the
corresponding value is much lower: often only about $10^{-5}-10^{-4}$ 
or even less of the total mass can be removed per time step if
fluctuations of the mass transfer rate by up to several orders of
magnitude are to be avoided.   

This paper is organized as follows: First, we discuss in
Sect.~\ref{Sec:Physics} the necessary input physics before we show in
Sect.~\ref{Sec:continuous} that the mass transfer rate in our model is
physically stable. Subsequently, in Sect.~\ref{Sec:discrete:explicit}
we prove mathematically that for a simple explicit algorithm the
time-discretized mass transfer rate becomes unstable if $\Delta t$ is
greater than a critical value. We also give an estimate of how many
time steps are required to calculate the binary evolution. Next, we
show in Sect.~\ref{Sec:discrete:Feigenbaum} that for increasing
\(\Delta t \) the mass transfer rate undergoes a Feigenbaum scenario
\citetext{\citealp{Feigenbaum78, Feigenbaum80}, for a more accessible
review see, e.g., \citealp{Thompson}} and finally becomes chaotic. In 
Sect.~\ref{Sec:discrete:ControllingChaos} we discuss how the onset of
chaos can be suppressed in terms of ``Controlling Chaos'' and we
present an implicit iteration scheme for the integration of the mass
transfer rate. Finally, in Sect.~\ref{Sec:implicit} we discuss 
a number of points which have to be considered for a practical
implementation.

\section{Input physics}
\label{Sec:Physics}

In this paper we use the same nomenclature as in \citet{Buening04},
hereafter called Paper I. For the mass transfer rate we use
\begin{equation}
\label{Eq:MTR}
\dot{M}_{2}=-\dot{M}_{0}\, \exp \left( \frac{\Delta R}{H_{\text {P}}}\right) ,
\end{equation}
where
\begin{equation}
\label{Eq:DeltaR}
\Delta R:=R_{2}-R_{\text {R,2}}
\end{equation}
denotes the difference between the radius $R_2$ and the Roche radius
$R_{\text{R},2}$ of the donor, $H_{\text{P}}$ the photospheric
pressure scale height, and $\dot{M}_0 > 0$ a weakly varying function
of several system parameters \citep[for details, see][]{Ritter88}.

Instead of (\ref{Eq:MTR}) other relations between $\dot{M}_{2}$ and
$\Delta R$ have also been used in the literature. For example
\citet{Tout97} have adopted a power-law dependence $\dot{M}_{2}
\propto (\Delta R/R_2)^3$. However, a mass transfer prescription other
than (\ref{Eq:MTR}) results mainly in a different characteristic scale
length
\begin{equation}
\label{Eq:ScaleLength}
H := \left( \Ableitung{\ln (-\dot{M}_2)}{\Delta R} \right)^{-1}
\end{equation}
at the secular mass transfer rate $\overline{\dot{M}}_2$ which is
given by 
\begin{equation}
\label{Eq:MTR:secular}
\tau_{\text{M}} := - \frac{M_2}{\,\overline{\dot{M}}_2}
= (\zeta_{\text{s}} - \zeta_{\text{R}}) \tau_{\text{d}}'.
\end{equation}
Here, $\tau_{\text{M}}$ is the secular time scale of the mass loss,
$\tau_{\text{d}}'$ the driving time scale including thermal relaxation
of the donor, $\zeta_{\text{s}}$ and $\zeta_{\text{R}}$ are
respectively the adiabatic mass radius exponent of the donor and
the mass radius exponent of the Roche radius, where $\zeta_\text{s} >
\zeta_\text{R}$ is required for physical stability of mass transfer  
(for details, see Paper I). For the mass transfer prescription
(\ref{Eq:MTR}), we have $H \equiv H_{\text{P}}$.

\section{The time-continuous system}
\label{Sec:continuous}

The mass transfer rate $\dot{M}_2$ is uniquely invertibly coupled to
$\Delta R$ by (\ref{Eq:MTR}). For simplicity, we will consider the
time evolution of $\Delta R$ instead of $\dot{M}_2$ which is given by
Eq.~[36] of Paper I:
\begin{equation}
\label{Eq:dDeltaRdt}
\Ableitung{}{t} \Delta R = (\zeta_{\text{s}} - \zeta_{\text{R}})~R_2
\frac{\dot{M}_2(\Delta R)}{M_2} + \frac{R_2}{\tau_{\text{d}}'}
=: F(\Delta R).
\end{equation}
Since for our stability considerations we neither take into account
changes in the system parameters ($\zeta_{\text{s}}$,
$\zeta_{\text{R}}, \tau_{\text{d}}'$) nor irradiation of the donor as
in Paper I, our model is completely described by this 1-dim.\
autonomous differential equation which simplifies the linear stability
analysis significantly. 

At the stationary value $\overline{\Delta R}$ which is the only fixed
point (FP) of (\ref{Eq:dDeltaRdt}) and which is equivalent to the
stationary mass transfer rate, using (\ref{Eq:MTR}) we get
\begin{equation}
\label{Eq:DF}
DF(\overline{\Delta R}) := \Ableitung{F(\overline{\Delta R})}{\Delta R}
=\left( \zeta_{\text{s}} - \zeta_{\text{R}} \right)
\frac{R_{2}}{H_{\text{P}}} \frac{\,\overline{\dot{M}}_{2}}{M_{2}}
\end{equation}
(cf. Eq.~[44] of Paper I). Since $\dot{M}_2 < 0$ and 
$\zeta_{\text{s}} > \zeta_{\text{R}}$, $DF$ is negative, not only for
$\overline{\Delta R}$, but even for all $\Delta R \in \R$. This means
that the stationary mass transfer rate, i.e., $\overline{\Delta R}$ is
stable and that all solutions $\Delta R(t)$ of (\ref{Eq:dDeltaRdt})
converge\footnote{%
This can be proven by using the fact that $F$ is continuous and has
only one FP. Thus, $DF(\Delta R) < 0$ for all $\Delta R \in \R$
together with $F(\overline{\Delta R}) = 0$ implies $F \gtrless 0
\Leftrightarrow \Delta R \lessgtr \overline{\Delta R}$
so that from (\ref{Eq:dDeltaRdt}) the convergence of all solutions
to $\overline{\Delta R}$ can be concluded.
}
to $\overline{\Delta R}$. The convergence occurs on a time scale of
\begin{equation}
\tau \approx \frac{H_{\text{P}}}{R_2} \tau_\text{d}'
\end{equation}
since 
\begin{equation}
 \Delta R(t) =       \overline{\Delta R} \pm
        \left[ \Delta R(0) - \overline{\Delta R} \right]\, {\rm e}^{-DF\,t}
\end{equation}
in the linearized system. This has also been discussed in more detail
by \citet{Antona89}.

\section{The time-discretized system}
\label{Sec:discrete}

\subsection{The explicit algorithm}
\label{Sec:discrete:explicit}

The simplest way to obtain the donor mass numerically as a function
of time is an explicit forward integration of $\dot{M}_2$ as given by
(\ref{Eq:Ansatz:explicit}). In this time-discretized system the
evolution of $\Delta R$ is given by an iteration equation which
follows directly from (\ref{Eq:dDeltaRdt}): 
\begin{eqnarray}
\label{Eq:Iteration:DeltaR}
\Delta R_{\text {n+1}} & = & \Delta R_{\text {n}}
+ \left[ \frac{R_{2}}{\tau_{\text {d}}'}
+ \left( \zeta _{\text{s}} - \zeta_{\text{R}} \right)
\frac{R_{2}}{M_{2}} \dot{M}_{2}(\Delta R_\text{n}) \right] \, \Delta t
\nonumber\\ 
& =: & \Phi \left( \Delta R_{\text{n}} \right) .
\end{eqnarray}
Here, $\Delta R_\text{n}$ denotes the value of $\Delta R$ at $t =
t_\text{n}$ which is in practice a (numerical) approximation for the
``real'' value of $\Delta R$ in the time-continuous system.

Obviously, (\ref{Eq:dDeltaRdt}) and (\ref{Eq:Iteration:DeltaR}),
i.e., the time-continuous and the time-discretized systems have the
same FP since $F(\Delta R) = 0$ if and only if $\Phi(\Delta R) =
\Delta R$. But, since $\Phi(\Delta R)$ is a map while $F(\Delta R)$ is
a vector field\footnote{%
For details on maps, vector fields an their related nomenclature, 
the reader is referred to, e.g., \citet{Guckenheimer}. 
}, 
the condition for stability is different: according to the
Hartmann-Grobmann theorem \citep[e.g.,][]{Guckenheimer}, a FP 
\( \vec{x} \) of a map \( \vec{\Phi } \) is stable if the absolute
values of all eigenvalues of the Jacobi matrix 
\( \vec{D}\vec{\Phi } \) of \( \vec{\Phi } \) at \( \vec{x} \) are
less than unity, and \( \vec{x} \) is unstable if at least one
eigenvalue has an absolute value greater than unity. 

In the case of $\Phi(\Delta R)$ from (\ref{Eq:Iteration:DeltaR}) this
means: $\overline{\Delta R}$ is stable if
\begin{equation}
\label{Eq:explicit:condition}
\left| D\Phi \right| := \left| \Ableitung{\Phi }{\Delta R}\right| 
= \left| 1+ \left( \zeta_\text{s} - \zeta_\text{R} \right)
\frac{R_{2}}{H_{\text{P}}}
\frac{\, \overline{\dot{M}}_{2}}{M_{2}} \Delta t\right| < 1,
\end{equation}
and unstable if the absolute value is greater than unity. As can be
easily shown by using (\ref{Eq:MTR:secular}), 
Eq.~(\ref{Eq:explicit:condition}) is equivalent to 
\begin{equation}
\label{Eq:explicit:Deltat}
\Delta t < 2 \frac{H_{\text{P}}}{R_{2}} \tau_{\text {d}}'
=: \Delta t_{\text{max}}.
\end{equation}
This means that in the neighbourhood of $\overline{\Delta R}$ all
solutions of the explicit iteration scheme (\ref{Eq:Ansatz:explicit})
converge to the FP if $\Delta t$ is less than the critical time step
length $\Delta t_{\text{max}}$. But for $\Delta t > \Delta
t_{\text{max}}$ the FP is unstable and the solutions of the 
time-discretized system (\ref{Eq:Iteration:DeltaR}) diverge although
the solutions of the time-continuous system (\ref{Eq:dDeltaRdt})
converge to the FP. 

If the initial value of the iteration is $\Delta R_0$, then in the
linearized system the orbit of $\Delta R_0$, i.e., $\{\Delta R_0,
\Delta R_1, \Delta R_2, \ldots \}$ is given by 
\begin{equation}
\label{Eq:Iteration:linearized}
\Delta R_{\text{n}+1} = \Phi(\Delta R_\text{n})
\approx \overline{\Delta R}
+ \left(D\Phi(\overline{\Delta R})\right)^\text{n+1}\, \left(\Delta
R_0 - \overline{\Delta R} \right),
\end{equation}
which can be shown by induction over $n$.

The following cases are possible (without proof): 
\begin{enumerate}
\item $\Delta t < \frac{H_\text{P}}{R_2} \tau_\text{d}'$: the orbit
      converges directly to the FP; for $\Delta t \rightarrow 0$ the
      continuous limit is reached.
\item $\Delta t = \frac{H_\text{P}}{R_2} \tau_\text{d}'$: the orbit
      converges after one iteration to the FP.
\item $\frac{H_\text{P}}{R_2} \tau_\text{d}' < \Delta t < 2
      \frac{H_\text{P}}{R_2} \tau_\text{d}'$: since
      $D\Phi(\overline{\Delta R})$ becomes negative, the orbit 
      converges alternatingly to the FP.
\item $2 \frac{H_\text{P}}{R_2} \tau_\text{d}' < \Delta t$: the orbit
      diverges.
\end{enumerate}

Therefore, any binary evolutionary code which uses an explicit
iteration scheme like (\ref{Eq:Ansatz:explicit}) encounters this
numerical instability. This is the case even if we assume that the
binary evolutionary code solves the equations of stellar structure
with infinite accurracy because this instability is a result of the 
time-discretization itself which is basically unavoidable for
numerical computations. 

It is now possible to estimate how many time steps an explicit
method like (\ref{Eq:Ansatz:explicit}) needs at least for the
computation of a mass transfer phase during which the maximum time
step length $\Delta t_\text{max}$ is used. From (\ref{Eq:MTR:secular})
and (\ref{Eq:explicit:Deltat}) it follows that at most the mass
fraction  
\begin{equation}
\label{Eq:explicit:DeltaM_max}
\frac{\Delta M_\text{max}}{M_2}
= \left| \frac{ \overline{\dot{M}_2} }{M_2}\right| \Delta t_\text{max}
= \frac{2}{\zeta_\text{s} - \zeta_\text{R}} \frac{H_\text{P}}{R_2} 
\end{equation}
can be removed from the donor per time step. After $n$ time steps the
initial mass $M_2^{(0)}$ has been reduced to 
\begin{equation}
\label{Eq:explicit:estimate}
M_2({t_\text{n}}) = \left(1 - \frac{2}{\zeta_\text{s} - \zeta_\text{R}}
\frac{H_\text{P}}{R_2} \right)^\text{n} M_2({t_0}).
\end{equation}
For low-mass main sequence (MS) stars, $\zeta_s \gtrsim -\frac{1}{3}$
\citep{Hjellming87, Hjellming-PHD}; for sufficiently small mass ratio
$M_2 / M_1$ of the binary, $\zeta_\text{R} \rightarrow -\frac{5}{3}$
in the analytical approximation of \citet{Paczynski71}, and therefore
$\zeta_s - \zeta_\text{R} \approx 1$. As an example: for  
$\frac{H_\text{P}}{R_2} = 10^{-4}$ which is a typical value for
low-mass MS stars and the low-mass limits $\zeta_s = -\frac{1}{3}$
and $\zeta_\text{R} = -\frac{5}{3}$ at least 4600 time steps are
necessary to reduce the donor mass by a factor of two according to
(\ref{Eq:explicit:estimate}). This is the most optimistic limit where
$\Delta t = \Delta t_\text{max}$. For thermally unstable systems,
which can even approach the onset of dynamical instability
\citetext{\citealp{Hjellming87}, \citealp{Schenker02}, and
\citealp{Podsiadlowski02}}, easily $10^4$ or $10^5$ time steps are
necessary to reduce the donor mass by a factor of two.

The main reason for this increased computational demand in the
case of thermally unstable mass transfer can be understood as follows:

The donor star of a binary in which mass transfer is thermally
unstable has a deep radiative envelope. Unperturbed stars with a deep
radiative envelope can be described (to some extent) by a polytropic
stellar structure with a polytropic index $n \lesssim 3$ and an
adiabatic index $\gamma = 5/3$ (monoatomic ideal gas), see e.g.
\cite{Hjellming87}, Table 1. Because for polytropes of index $n$,  
$\zeta_s = (1-n)/(3-n)$, for radiative stars (with $n \lesssim 3$)
$\zeta_s$ is a very large, positive number. However, this holds only
for unperturbed stars in thermal equilibrium and only in the limit of
infinitesimally small mass loss (see \cite{Hjellming87}, last
paragraph of Sect. II). On the other hand, for finite mass loss
$\zeta_s$ decreases rapidly with the amount of mass lost. Yet for such
stars, at least during the initial phases of thermal timescale mass
transfer, $\zeta_s$ is still much larger than unity, i.e. typically of
order $10 - 100$ (as shown in \cite{Hjellming87}, their Figs. 3 and
4). As a consequence, $\zeta_s - \zeta_\text{R}$ is then also of the
order of $10 - 100$, and $\Delta M_\text{max}$ as given in Eq.
(\ref{Eq:explicit:DeltaM_max}) is smaller and the minimum number of
required time steps increases by that factor.

As can be seen from (\ref{Eq:explicit:estimate}), an artificial
increase of $H_\text{P}$ can significantly reduce the  required 
number of time steps. This has been used in the past by numerous
authors to speed up the computations \citep[e.g.,][]{Hameury91}.
This approach yields the correct mass transfer rate only when mass
transfer is close to stationary. However, it cannot be used if the
turn-on and turn-off of mass transfer is important, as is the case
for irradiation-induced mass transfer cycles (cf. Paper I).

\subsection{The Feigenbaum scenario}
\label{Sec:discrete:Feigenbaum}

The linear stability analysis discussed in
Sect.~\ref{Sec:discrete:explicit} describes only the system dynamics
near the FP, i.e., the local dynamics; we will now briefly discuss the
global dynamics.

When going to dimensionless quantities
\begin{equation}
\label{Eq:scaling}
x := \frac{\Delta R - \overline{\Delta R}}{H_\text{P}},\quad
\delta := \frac{R_2}{H_\text{P}} \frac{\Delta t}{\tau_\text{d}'}
\end{equation}
and using (\ref{Eq:MTR}) and (\ref{Eq:MTR:secular}),
Eq.~(\ref{Eq:Iteration:DeltaR}) is equivalent to
\begin{equation}
\label{Eq:Ansatz:dimensionless}
x_{\text{n}+1} = x_\text{n}
+ \left[ 1 - \exp(x_\text{n}) \right] \,\delta
=: x_\text{n} + G(x_\text{n})
=: f_\delta(x_\text{n}).
\end{equation}
The family of functions $f_\delta$ is topologically conjugated to a
family of $\cal S$-unimodal functions \footnote{%
Unimodal functions and their system dynamics are discussed in detail
by \citet{Collet80}. For an overview about topological conjugacy and
the Feigenbaum scenario see, e.g., \citet{Thompson, Jackson89}.
For some background about topological conjugacy and topological
equivalence of maps, see \citet{Arnold83, Wiggins_Dynamical}. A more
in-depth discussion of map (\ref{Eq:Ansatz:dimensionless}) can also
be found in \citet{Buening-PHD}.
}: Thus, $f_\delta$ is qualitatively similar to the well-known
logistic map: 
\begin{equation}
\label{Eq:LogisticMap}
g_\delta(x) = \delta x\,(1 - x)
\end{equation}
and undergoes a Feigenbaum scenario for increasing $\delta$ which is
called the control or chaos parameter\footnote{%
$f_\delta$ is not topologically conjugated to a full family
of $\cal S$-unimodal functions. Hence, unlike the logistic map,
$f_\delta$ formally does not undergo a complete Feigenbaum scenario.
}.

\begin{figure}[ht]
\includegraphics[width=0.5\textwidth]{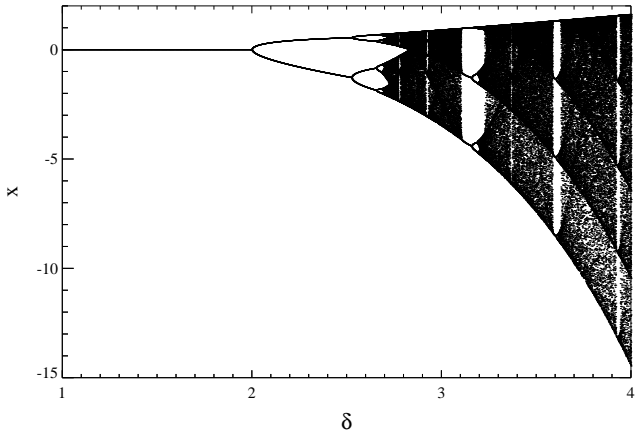}
\caption{Feigenbaum diagram of map (\ref{Eq:Ansatz:dimensionless}).
50 iterations of one orbit for 5000 different values of $\delta$
are shown.
\label{Fig:theory}}
\end{figure}

For $\delta = 2$, corresponding to $\Delta t = \Delta t_\text{max}$,
the FP becomes unstable and bifurcates into an unstable FP and a stable
cycle of period 2. For increasing values of $\delta$, the 2-cycle also
becomes unstable, and a stable 4-cycle appears, and so on. At a
critical value of $\delta \approx 2.7$, the system dynamics become 
chaotic, and beyond this value the chaotic dynamics are permanently
interrupted by finite windows of regular dynamics which correspond to
the existence of a stable cycle of any period. Figure~\ref{Fig:theory}
shows $50$ iterations of (\ref{Eq:Ansatz:dimensionless}) of one single
orbit for $5000$ different values of $\delta$. The FP $\bar{x} 
\equiv 0$  corresponds to $\overline{\Delta R}$. At about $\delta
\approx 3.1$, a stable orbit of period $3$ appears, whose existence is
a mathematical proof for the existence of chaotic dynamics in the
system according to the famous ``Period three implies chaos'' by
\citet{Yorke75}. $x$ is a logarithm of $\dot M_2$ and $\delta$ is
linear in $\Delta t$. Thus, the transitional region from stable to
chaotic dynamics is rather small.

Since chaotic or ``random'' values in the mass transfer rate are not
desirable, how can the chaotic dynamics be suppressed? As mentioned
above, an artificial increase of $H_\text{P}$ will increase
$\Delta t_\text{max}$ and therefore shift the onset of chaos to higher
values of $\delta$. In general, the choice of a mass transfer
prescription which differs from (\ref{Eq:MTR}) will only shift the
onset of chaos to different values of $\delta$ but it cannot suppress
it. At least, if $\dot{M}_2$ grows as $x^\alpha$, $f_\delta$
basically keeps its behaviour (without proof) for $\alpha \geq 2$ and
this is the case for all physical models for computing mass transfer.

\subsection{Controlling Chaos}
\label{Sec:discrete:ControllingChaos}

It is possible to suppress chaotic dynamics and to stabilize a given
FP by introducing small pertubations in terms of ``Controlling Chaos''
\footnote{%
The term ``Controlling Chaos'' comes from an article of the same title
by \cite{Grebogi} who suggested to control chaotic motion in nonlinear
systems by small perturbations. The keyword ``Controlling Chaos'' or
``Control of Chaos'' later became common in that special field of
research. In their introduction to {\it The Control of Chaos: Theory
and Applications}, \cite{Boccaletti00} gave a short historical
overview of that topic.
}. 
There are two distinct approaches: the first one varies one system
parameter by a small amount $\varepsilon_\text{n}$ for each time step
to  enforce stability of the FP \citep{Grebogi}, but this method
requires the a priori knowledge of the FP, and this is not the case
for our system. The second method adds a small correction 
$\varepsilon_\text{n}$ to the new iteration value, i.e.,
\begin{equation}
\label{Eq:ControllingChaos}
x_{\text{n}+1} := f(x_\text{n}) + \varepsilon_\text{n}.
\end{equation}
The simplest approach uses 
$\varepsilon_\text{n} = K\,(x_\text{n} - x_\text{n-1})$
with a suitable constant $K$ \citep{Pyragas92}. This so-called delayed
dynamical feedback involves $x_\text{n-1}$ and $x_\text{n}$, i.e., it
uses information from previous iteration steps. This can extend the
stability limit to larger values of $\delta$ depending on $K$, but as
numerical experience shows, this is by far not sufficient in our case.
The inclusion of \emph{all} prior iteration values \citep{Socolar94}
like $\varepsilon_\text{n} = K\,(x_\text{n} - x_\text{n-1}) + 
R \varepsilon_\text{n-1}$ with a suitable constant $R$ achieves
significantly better results. For $R\rightarrow 1$ the stability of
the FP can be extended to arbitrarily large values of $\delta$ but at
the expense of an arbitrarily small basin of attraction which makes
this approch difficult for practical application. 

A nonlinear approach of the form 
\begin{equation}
\label{Eq:Ansatz:Lichtenberg}
x_\text{n+1} = f(x_\text{n-1}) + K\,(f(x_\text{n}) - f(x_\text{n-1}))
+ R \varepsilon_\text{n}
\end{equation}
as has been proposed by \citet{Lichtenberg96} yields a significant
improvement of the basin of attraction. For the specific choice of
$R=K$ and $K = (Df(\bar{x}) - 1)^{-1} DF(\bar{x})$, the FP becomes
superstable, i.e., the FP iteration (\ref{Eq:Ansatz:Lichtenberg})
converges quadratically in a sufficiently small neighbourhood of the
FP.

Since our map (\ref{Eq:Ansatz:dimensionless}) is of the form 
$f_\delta(x) = x + G_\delta(x)$, Eq.~(\ref{Eq:Ansatz:Lichtenberg}) 
is equivalent to 
\begin{equation}
\label{Eq:Ansatz:Newton}
x_\text{n+1} = x_\text{n} + DG_\delta^{-1}(\bar{x}) G_\delta(x_\text{n}).
\end{equation}
Because the position of the FP is a priori unknown, the best available
approximation to $DG_\delta(\bar{x})$ is given by
$DG_\delta(x_\text{n})$. Then (\ref{Eq:Ansatz:Newton}) turns into a 
standard Newton's method for $G_\delta(x)$. Therefore, we conclude
that Newton's method is most likely the best available method to
compute $\bar{x}$, and $\overline{\dot{M}_2}$ with a reasonable
computational effort.

\section{The implicit algorithm}
\label{Sec:implicit}

While the explicit iteration scheme is a simple time integration
scheme which requires \emph{one} iteration per time step, the proposed
implicit iteration scheme is a fixed point search which requires
\emph{several} iterations per time step until the fixed point is found
with sufficiently high accuracy. 

Fortunately, the equations of stellar structure are typically solved
by Newton's method, more explicitly by the so-called Henyey method
\citep{Hofmeister, Kippenhahn}. Thus, the best solution is  to include
the mass transfer rate or the stellar mass itself as an additional
variable in the Henyey method and to perform the fixed point search
simultaneously with the solution of the equations of stellar structure
(Paper I). This has also been done independently by \citet{Benvenuto03}. 
For details of our implementation, see \citet{Buening-PHD}.

\begin{figure}[ht]
\includegraphics[width=0.5\textwidth]{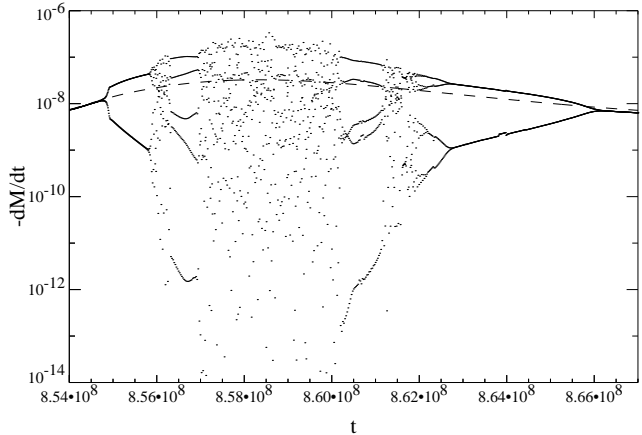}
\caption{Mass transfer rate $\dot{M}_2 [M_\sol / \text{yr}]$ as 
a function of time $t[\text{yr}]$ for an early case A mass transfer
in a binary system with $M_1 = M_2 = 2 M_\sol$ and using a constant
time step length of $5000\,\text{yr}$. The variation of the chaos
parameter $\delta$ is caused by the variation of thermal relaxation.
The system dynamics are most unstable when $\tau_\text{d}'$ is minimal.
Every dot corresponds to one single time step in the calculation with
the explicit scheme (\ref{Eq:Ansatz:explicit}); the dashed line shows
the result of the same computation, but using the implicit method.
\label{Fig:practise}
}
\end{figure}

Figure~\ref{Fig:practise} shows the mass transfer rate obtained with
our binary evolutionary code as a function of time for one specific
binary system. Since our evolutionary code has not been designed for
the analysis of chaotic dynamics, we have kept $\Delta t$ constant and
instead have used the fact that the timescale of thermal relaxation
varies, which is the dominant term in $\tau_\text{d}'$ for the binary
system in question. When mass transfer starts, the thermal relaxation
of the donor increases (i.e., $\tau_\text{d}'$ decreases, c.f.
Eq.~(\ref{Eq:scaling})), and, therefore, $\delta$ also increases. At
some point, thermal relaxation reaches a maximum and decreases
afterwards, and so does $\delta$.

When using the explicit algorithm (\ref{Eq:Ansatz:explicit}), the
resulting mass transfer rate undergoes a period doubling bifurcation
before it exhibits chaotic dynamics. Within the chaotic region a
periodic window of period 5 appears. After thermal relaxation and
hence also $\delta$ has reached its maximum, the Feigenbaum scenario
evolves backwards. Since the decrease of $\delta$ occurs slower than 
its increase before, even stable cycles of period 8 and 4 appear
before the secular mass transfer rate finally becomes stable. In
contrast, the result obtained with our implicit algorithm for the 
same system and the same system parameters is shown by the dashed
line.

Although the implicit algorithm stabilizes the FP and even provides a
quadratic convergence of the iteration, chaotic dynamics are still
present outside of a neighbourhood of the FP. Therefore, a good
initial value for the iteration is necessary. We used a linear
extrapolation of $\dot{M}_0$ and $H_\text{P}$ to determine the initial
value for $\dot{M}_2$ by (\ref{Eq:MTR}). First, we perform typically
2-4 iterations and keep the preestimated value for $\dot{M}_2$
constant until the solution for the other stellar parameters has
almost converged. Otherwise, even small fluctuations of $R_2$ might
push the next iteration value of $\dot{M}_2$ out of the basin of
attraction of the FP and prevent convergence. Then, we finish with
mostly 2-6 iterations using the full implicit algorithm to determine
the correct mass transfer rate.

Furthermore, to calculate the mass transfer rate with a numerical
accuracy of better than $1\%$, the stellar radius has to be determined
with a very high numerical accurracy. According to (\ref{Eq:MTR}),
for a scale height of {$H_\text{P} \approx 10^{-4}\, R_2$, which is
typical for low-mass MS stars, $R_2$ has to be determined with a
relative accuracy of the order of $10^{-6}$ in order to get
$\dot{M}_2$ with a relative accuracy of the order of $10^{-2}$. To
compute the radius with such a high accuracy the stellar physics and
especially the equation of state must be a very smooth function of its
variables. Every small discontinuity, especially in the outer layers 
of the star, can cause small jumps in the stellar radius which appear,
magnified by the factor $R_2/H_\text{P}$, as significant jumps in the
mass transfer rate.

\section{Summary and conclusions}
\label{Sec:summary}

We have used a simple analytical model, a 1d autonomous ordinary
differential equation to describe the time evolution of the mass
transfer rate. We have shown that, while the FP in the time-continuous
system, i.e., the ``physical'' mass transfer rate is stable, the FP in
the time-discretized system, i.e., the ``numerical'' mass transfer
rate becomes unstable if the length of the time step $\Delta t$
exceeds a critical value $\Delta t_\text{max}$ given by
(\ref{Eq:explicit:Deltat}). We have estimated that even in the ideal
case where $\Delta t = \Delta t_\text{max}$ at least several thousand
time steps are necessary to reduce the donor mass in a low-mass binary
system by a factor of two. For systems with thermally unstable mass
transfer, it is even worse. 

We outline a mathematical proof that the iteration equation for the 
time-discretized system shows a behaviour similar to that of the
logistic map and, for $\Delta t > \Delta t_\text{max}$, undergoes a
series of period doublings which finally leads to chaotic dynamics,
i.e., to apparently random values of the computed mass transfer rate.
The choice of a different explicit prescription to calculate the 
mass transfer rate results only in a shift of the critical time step
length $\Delta t_\text{max}$ which, in turn, depends on the
characteristic scale length $H$ at the FP $\overline{\Delta R}$,
i.e., at the secular mass transfer rate.

In terms of ``Controlling Chaos'' we have  briefly discussed several 
methods to stabilize the FP. Various modified iteration schemes which
are equivalent to different explicit iteration schemes have been 
discussed in the literature, but they all do not show sufficient 
stabilization. Therefore, we suggest that using a different explicit 
iteration scheme may shift $\Delta t_\text{max}$ to higher values but
will not solve the problem. An implicit scheme, Newton's method, is
the most promising solution because it stabilizes the FP formally for
$\Delta t \rightarrow \infty$, although for this to be the case a
sufficiently good initial value for the iteration is required. 

In practice, the implicit algorithm reduces the number of required
time steps by at least a factor of 10. Another advantage is that the
implicit algorithm either yields the ``correct'' mass transfer rate or
does not converge at all, whereas the explicit algorithm provides a
result in every case, even if it is random. Therefore, in our binary
evolutionary calculations we reject results of the last time step if
convergence is not reached and restart it with a smaller $\Delta t$.

\begin{acknowledgement}
We thank A. Weiss and H. Schlattl for providing their stellar
evolutionary code, and H. Schlattl for valuable support and helpful
discussions about numerics. We also thank U. Kolb for providing 
unpublished details about his numerical calculations. 

\end{acknowledgement}

\bibliographystyle{aa}
\bibliography{astro,klassik}

\end{document}